\documentclass[conference,a4paper]{IEEEtran}
\usepackage{eso-pic}

\usepackage{graphicx}

\begin{document}
\title{A Cross-Layer Approach for Video Delivery over Wireless Video Sensor Networks}

%\author{Othmane Alaoui-Fdili$^{1,2}$, Fran\c{c}ois-Xavier Coudoux$^2$, Youssef Fakhri$^{1,3}$, Patrick Corlay$^2$, Driss Aboutajdine$^1$.}
%\address{$^1$LRIT-CNRST URAC29, Universit\'e Mohammed V-Agdal, Rabat, Morocco;\\ $^2$ IEMN UMR 8520, Department OAE, UVHC, Valenciennes, France; \\ $^3$ LARIT \'equipe r\'eseaux et T\'el\'ecommunications, Universit\'e IbnTofail, K\'enitra, Morocco. }


%\author{The order of authors: ICIP or EUSIPCO or any particular preference?}
\author{\IEEEauthorblockN{Othmane Alaoui-Fdili\IEEEauthorrefmark{1}\IEEEauthorrefmark{4},
Patrick Corlay\IEEEauthorrefmark{4},
Youssef Fakhri\IEEEauthorrefmark{1}\IEEEauthorrefmark{3},
Fran\c{c}ois-Xavier Coudoux\IEEEauthorrefmark{4} and
Driss Aboutajdine\IEEEauthorrefmark{1}}
\IEEEauthorblockA{\IEEEauthorrefmark{1}LRIT-CNRST URAC29, Universit\'e Mohammed V-Agdal, Rabat, Morocco}
\IEEEauthorblockA{\IEEEauthorrefmark{4}IEMN UMR 8520, Department OAE, UVHC, Valenciennes, France}
\IEEEauthorblockA{\IEEEauthorrefmark{3}LARIT \'equipe r\'eseaux et T\'el\'ecommunications, Universit\'e IbnTofail, K\'enitra, Morocco}
}


%\AddToShipoutPicture*{\small \sffamily\raisebox{1.8cm}{\hspace{1.8cm}978-1-4244-5997-1/10/\$26.00 \copyright2010 IEEE}}

\maketitle

\begin{abstract}
In this paper, we propose a novel cross-layer approach for video delivery over Wireless Video Sensor Networks (WVSN)s.  We adopt an energy efficient and adaptive video compression scheme dedicated to the WVSNs, based on the H.264/AVC video compression standard. The encoder operates using two modes.  In the first mode, the nodes capture the scene following a low frame rate. When an event is detected, the encoder switches to the second mode with a higher frame rate and outputs two different types of macroblocks, referring to the region of interest and the background respectively. Furthermore, we propose an Energy and Queue Buffer Size Aware MMSPEED-based protocol for reliably and energy efficiently routing both regions towards the destination. Simulations results prove that the proposed approach is energy efficient and delivers good quality video streams. In addition, the proposed routing protocol EQBSA-MMSPEED outperforms its predecessors, the QBSA-MMSPEED and the MMSPEED, providing 33\% of lifetime extension and 3 dBs of video quality enhancement.
%\boldmath
%In this paper, we propose a novel cross-layer approach for video delivery over the Wireless Video Sensor Networks (WVSN)s.  We adopt an energy efficient and adaptive video compression scheme dedicated to WVSN, based on the H.264/AVC standard. The encoder operates using two modes, namely the Standby mode and the Rush mode. In the first mode, the nodes capture the scene following a low frame rate. When an event is detected, a higher frame rate is used and the encoder outputs two different types of macroblocks referring to the region of interest and the background respectively. For energy efficiency, the global bit rate is adaptively adjusted to each region. Then, we propose an Energy and Queue Buffer Size Aware MMSPEED-based (EQBSA-MMSPEED) protocol for routing  both regions to the destination. Simulations prove that the proposed approach is energy efficient and delivers high quality video streams. In addition, the proposed routing protocol EQBSA-MMSPEED outperforms its predecessors, the QBSA-MMSPEED and the MMSPEED,  in terms of reliability as well as energy.
\end{abstract}


% \fi
\IEEEpeerreviewmaketitle

\section{Introduction}
With the recent advances in image and video processing, wireless sensor nodes are doted of new capabilities that enable them the capture and the processing of visual information. Networks of such interconnected devices are called Wireless Video Sensor Networks (WVSN)s. They are actually investigated for various monitoring applications, for indoor and outdoor environments \cite{SurvWVSN2014}. In the WVSN, the video nodes collaborate in order to deliver visual information about an area on interest, to a destination called the sink, via multihops short range transmissions. The nodes operate with respect to the available and limited resources. In fact, these units are battery powered. The management of this component is therefore a crucial issue in the Wireless Sensor Network (WSN) context in general, and in the WVSN context in particular. In brief, the energy consumption has to be efficiently managed in all of the protocol stack's layers.

In the WVSNs, the video nodes compress the video streams prior to the transmission. Several works have been conducted in order to propose video compression schemes adapted to the WVSNs. In some works \cite{WMSNTranMPEG}-\cite{cross1}, the authors use a video compression standard as is and try to enhance the energy consumption in lower layers. While others \cite{aghdasi}-\cite{MDCWVSN} prefer to slightly modify the standard to make it more appropriate to the video nodes.

In \cite{fdili2013energy}, we adopt the second choice and propose an energy efficient and adaptive video compression scheme dedicated to the WVSNs. Actually this scheme relies on the H.264/AVC video compression standard \cite{Richardson} in its intra-only mode.  In addition it outputs two macroblocks categories depending on  which region each one belongs, namely the Region Of Interest (ROI) or the Background (BKGD). This is done to introduce the differentiated service paradigm that is strongly recommended for WVSN. The simulations results have proven the energy efficiency of the proposed scheme.

Consequently, for efficiently transferring the output streams, the underlying routing protocol has to be able to manage and serve different classes of packets at the same time. The authors in \cite{mmspeed} propose a Multipath Multi-SPEED protocol (MMSPEED) that is able to handle multiple traffic classes. Actually, it considers two quality domains, namely the timeliness and the reliability. In the first one, MMSPEED tries to maintain a given packet's Progression Speed (PS) across the network in order to meet the desired delay. Thereafter, to offer a Desired Reliability (DR), a Total Reaching Probability (TRP) is computed by injecting the Reaching Probability (RP) of each of the candidate nodes until the TRP reaches the DR. 

In \cite{our}, we investigate the MMSPEED protocol and show the gain that can be observed in timeliness and reliability domains when considering the Available Buffer Size (ABS) in the node's queue during the routing process. In fact, taking into account the ABS of neighbour nodes handles the congestion and leads to a decrease of the packets' experienced delay as well as the ratio of the dropped packets. Nevertheless, there is still a need to the consideration of the energy during the routing process.

Therefore, in this paper, we first present a cross-layer approach for video delivery over WVSNs based on the video compression scheme presented in \cite{fdili2013energy} and the routing protocol presented in \cite{our}. Then, we propose an Energy and Queue Buffer Size Aware MMSPEED (EQBSA-MMSPEED) protocol that introduces the residual energy as new routing metric in order to decrease the energy consumption and hence extending the network lifetime. In addition, EQBSA-MMSPEED proposes a way to enhance the packet delivery ratio, called the Last Chance Procedure (LCP).

The rest of this paper is organized as follows. In section \ref{app} the main contributions of this paper are in detail explained. Section \ref{Sim}, presents and discusses the simulations results in terms of energy consumption, delay, reliability and received video quality. Finally, section \ref{conclu} concludes the paper.
 
%\hfill January 11, 2007

\section{Proposed Approach}
\label{app}
In this Section we present in detail the proposed approach which is based on an energy efficient video compression scheme and an Energy and Queue Buffer Size Aware routing protocol.
\subsection{Energy Efficient Adaptive Video Compression Scheme}
\label{videocomp}
\begin{figure}[htb]
\includegraphics[width=8.75cm,height=3.5cm]{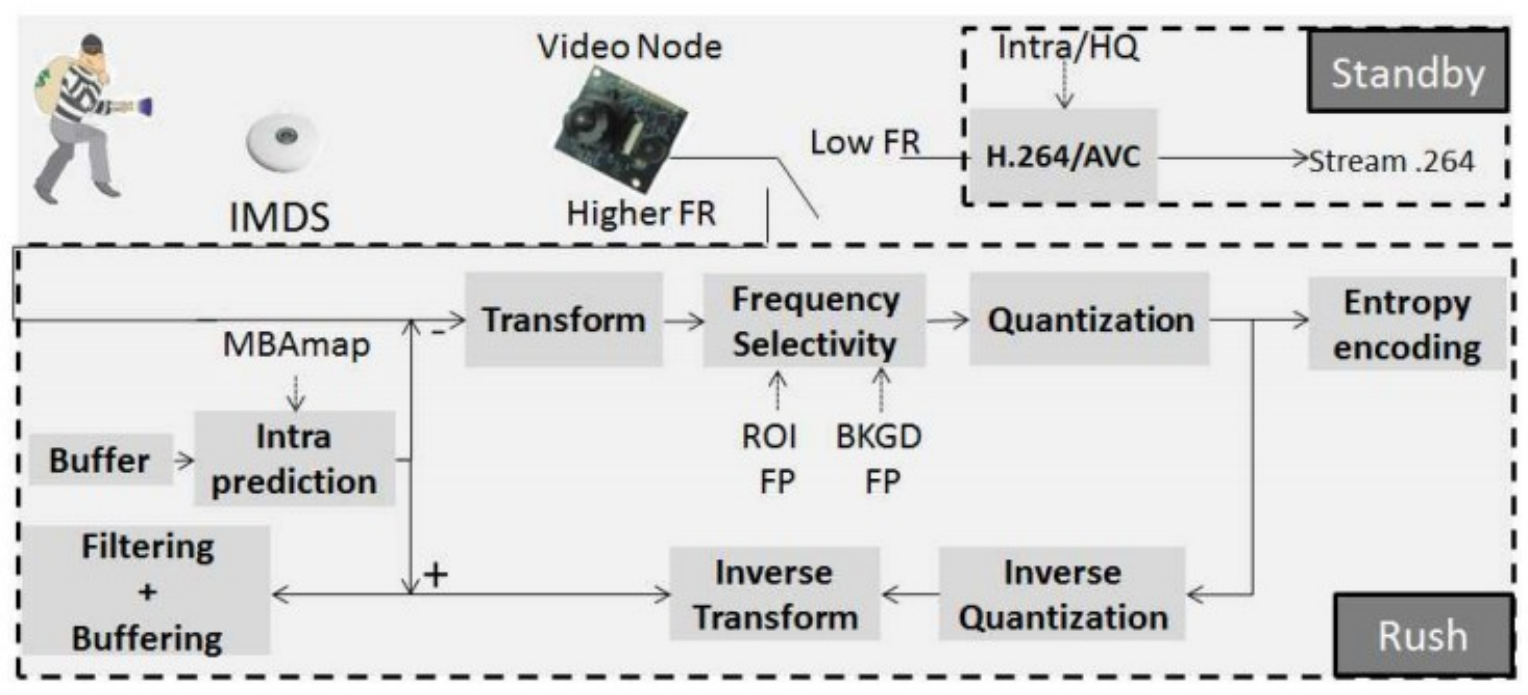}
\caption{Block diagram of the proposed scheme}
\label{sec3}
\end{figure}

 As shown on Fig.\ref{sec3} , the video compression scheme operates using two modes: the Standby and the Rush modes. In the first mode, the nodes capture the scene following a low Frame Rate (FR) to preserve their energies. Then, they compress the video signal with a given Quantization parameter (QP). In addition, they use the intra-only mode that has proven its efficiency and suitability to the WVSN context. In fact, the authors in \cite{JpegJ2K} propose a comparative study concerning the H.264/AVC intra mode against JPEG and JPEG2000. They conclude that the H.264/AVC intra mode offers an interesting compromise in terms of complexity, quality and coding efficiency. 

When an event occurs, the concerned nodes switch to the Rush mode. First, they adopt a higher frame rate to report the event. Then, the Flexible Macroblock Ordering (FMO) tool is used to produce two service differentiated streams corresponding to the ROI and the BKGD. Moreover, the nodes keep the same QP and apply a Frequency Selectivity (FS) bit rate adaptation in order to decrease the energy consumed for the transmission. Actually it consists on keeping a number of the coefficients, FP, after the transform operation. Note that the FS is a logical operation that doesn't add any complexity to the encoder, unlike the case when a requantization is used. Finally, the residual coefficients are entropy coded by the Context-Adaptive Variable-Length Coding (CAVLC) that can approach the entropy of the source with a reduced complexity, compared to the Context-Based Adaptive Binary Arithmetic Coding (CABAC). 
\subsection{Energy and Queue Buffer Size Aware MMSPEED Routing Protocol (EQBSA-MMSPEED)}
 Now that the video streams are encoded in two different categories, we need to transfer them to the sink.  For this purpose, a service differentiated routing protocol is needed to serve each flow according to its traffic class (ROI or BKGD). In addition, this routing protocol has to consider the compulsory energy constraint,  especially when dealing with energy consuming streams. 

First of all, let us mathematically define some terms that will be used. The first one is the Forwarding Set (FS), regrouping the nodes that are in the communication range and closer to the destination $D$ than the current node $i$, and it is formulated as follows:
\begin{equation}
FS_i(D)= \left\lbrace j \in NS_i /dist(i,D)-dist(j,D)>0 \right\rbrace 
\end{equation}
where $dist(i,D)$ and $dist(j,D)$ are the Euclidean distances between $i$ and $D$, $j$ and $D$ respectively, and $NS_i$ is the neighbourhood set that contains the one-hop reachable neighbours. The Progression Speed (PS) of a packet towards the destination D, if the node $i$ forwards it to the node $j$ is given as follows \cite{mmspeed} :
 \begin{equation}
\label{speedeq}
Speed_{ij}(D)=\frac{dist(i,D)-dist(j,D)}{Delay_{ij}}
\end{equation} 
The Reaching Probability (RP) of a packet to the destination D, if the node $i$ forwards it to the node $j$ is given by \cite{mmspeed}: 
\begin{equation}
\label{RP}
RP^{D}_{ij}=(1-e_{ij})(1-e_{ij})^{[\frac{dist(j,D)}{dist(i,j)}]}
\end{equation}
where $e_{ij}$ corresponds to the rate of lost packets sent to $j$. Then, the Total Reaching Probability (TRP) is given by \cite{mmspeed}:
\begin{equation}
\label{TRP}
TRP_{New}=1-(1-TRP_{Old})(1-RP^{D}_{ij})
\end{equation} 

\begin{figure}[htb]
\centering \includegraphics[width=7cm,height=0.75cm]{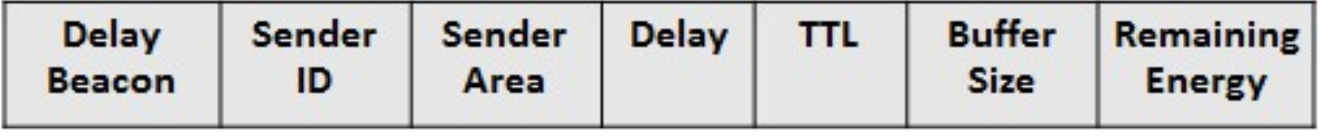}
\caption{Delay beacon structure in EQBSA-MMSPEED}
\label{packet}
\end{figure}

In order to consider the energy consumption during the routing process, we propose to inject the information about the residual energy in the delay packets that are periodically broadcasted by each node in its neighbourhood. Thus, the delay beacon in the proposed EQBSA-MMSPEED protocol has the following structure, reported by Fig \ref{packet}. 

\begin{figure}[htb]
\centering \includegraphics[width=7cm,height=1cm]{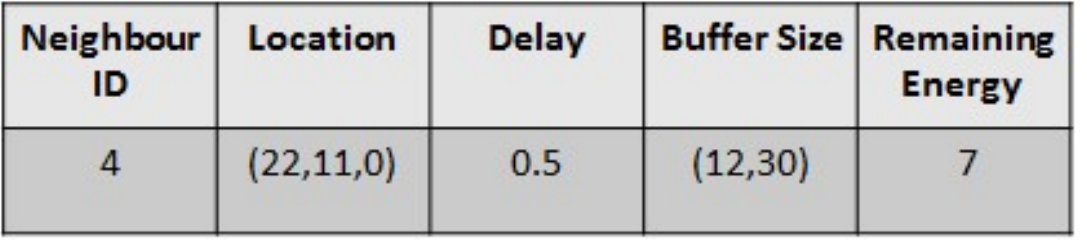}
\caption{Neighbour Table in EQBSA-MMSPEED}
\label{NT}
\end{figure}

Consequently, when  a node receives a delay packet, it updates the different fields of its neighbouring table that now contains the field of the remaining energy (see Fig.\ref{NT}).

When a node receives a data packet of traffic class $l$ to be routed, the node processes it as follows. The proposed EQBSA-MMSPEED protocol starts by constructing the FS, then the subsets $FS_{high}$ and $FS_{low}$ that contains respectively the nodes with PS higher and lower than the requested one. Afterwards, a score is assigned to each of the nodes of the $FS_{high}$ subset. Actually, in the EQBSA-MMSPEED protocol, this score is based on three metrics: the Residual Energy (RE), the Available Buffer Size (ABS) and the RP. The calculation of a neighbour node $j$'s score is given by: 
\begin{equation}
\label{eij}
score(j)=\alpha \times RP^{D}_{ij}+ \beta \times ABS_l(j) + (1-\alpha-\beta) \times RE(j)
\end{equation}
where $\alpha$ and $\beta$ are weighting coefficients to be fixed, and $ABS_l$ is the information about the available buffer space in node $j$'s queue at level $l$. Once the nodes are sorted according to the calculated scores, they are presented to the equation \ref{TRP} one by one until the TRP reaches the DR for the packet's traffic class. 

When the node fails to find forwarding nodes belonging to $FS_{high}$, MMSPEED protocol probabilistically drops the packet or forwards it to the best node in the $FS_{low}$ in terms of the PS. Therefore, in the proposed EQBSA-MMSPEED protocol, a new procedure called the Last Chance Procedure (LCP), is introduced. The purpose of this procedure is doing more network lifetime extension and achieving proper reliability even in such cases.
The LCP consists in first dividing the subset  $FS_{low}$ in two subsets, namely $FS_{low}^{LC}$ and $FS_{low}^{Rescue}$. Formally, $FS_{low}^{LC}$ can be defined as follows:
\begin{equation}
FS_{low}^{LC}=\left\lbrace j \in FS_{low} / PS(j)\geq PS_{avg} \right\rbrace
\end{equation} 
Hence, the $FS_{low}^{LC}$ contains the nodes of the $FS_{low}$ that are able to offer a packet's PS higher than the average PS in $FS_{low}$. Otherwise, the node belongs to $FS_{low}^{Rescue}$. Subsequently, the nodes of $FS_{low}^{LC}$ are sorted according to their scores, calculated by the equation \ref{eij}. Then, they are injected in the equation \ref{TRP} until the TRP reaches the DR for the packet's traffic class. 
Finally, in case the presented nodes are not sufficient to meet the DR, the best nodes in terms of reliability in $FS_{low}^{Rescue}$ are selected one by one.

In summary, the proposed EQBSA-MMSPEED protocol compared to its predecessors, is able to (1) meet the required deadlines since the nodes are chosen -when possible- from the $FS_{high}$, (2) to extend the network lifetime by taking into account the residual energy of the nodes and (3) tries to offer an acceptable packet delivery ratio as much as possible thanks to the LCP.  
\section{Simulations Results}
\label{Sim}
In order to validate the proposed approach, we conducted several simulations using the JSIM \cite{jsim1} simulator. We considered 100 nodes uniformly deployed in an area of interest where 50\% of them are video nodes. 
First, the nodes starts exchanging control packets for a given duration $XD$. Then  the video nodes start the Standby mode, capturing the scene following a low FR, $FR_{SM}$, and compress it using a given $QP$. Each frame is then subdivided into a number of packets to be routed towards the sink according to a given packet rate, $PR_{SM}$.
At the event time, the node that has detected it switches to the Rush mode following a higher FR , $FR_{RM}$. Each of the macroblocks of the captured frames is categorized as belonging to ROI or BKGD thanks to the FMO option. The BKGD component is then bit rate adapted using the FS of parameter $FP$. Afterwards, the ROI and the BKGD are encapsulated into packets to be routed towards the sink according to a  packet rate, $PR_{RM}$, with suitable reliability and delay constraints. 

For reliable evaluation, the source nodes are randomly designated in each realisation. We evaluate the performances of the proposed approach during the video transmission until the last deliverable packet reaches the sink. Table \ref{config} reports the used simulations parameters.

\begin{table}[htb]
\centering
\caption{Simulations setup}
\label{config}
\begin{footnotesize}
\begin{tabular}{l l}
\hline
\hspace{1.5cm} \textbf{Environment} & \hspace{-1.25cm} \textbf{Settings} \\
\hline
Video format & QCIF(176x144)\\
\hline
$FR_{SM}$ & 1 (fps)\\
\hline
$FR_{RM}$ & 3 (fps)\\
\hline
$QP$ & 32\\
\hline
$FP$ & 6\\
\hline
ROI ratio & 0.5\\
\hline
$XD$ & 50 (s)\\
\hline
 $PR_{SM}$ & 5 (pps)\\
\hline
 $PR_{RM}$ & 10 (pps)\\
\hline
Packets per Frame & 33\\
\hline
Queue size & 100 (Pckts)\\
\hline
QBSA-MMSPEED$(\alpha)$ & 0.7\\
\hline
EQBSA-MMSPEED$(\alpha,\beta)$ & (0.3,0.2)\\
\hline
$DR_{ROI}$ & 0.7\\
\hline
$DR_{BKGD}$ & 0.3\\
\hline
$Delay_{ROI}$ & 1 (s)\\
\hline
$Delay_{BKGD}$ & 2 (s)\\
\hline
Radio range & 40 (m)\\
\hline
Terrain & 200 x 200 (m$^2$) \\
\hline
Bandwidth & 250 (kbps)\\
\hline
\hspace{2cm} \textbf{Energy} & \hspace{-1.75cm} \textbf{Model} \\
\hline
Initial energy & 10 (Joules)\\
\hline
Current consumption for transmitting & 28.18 (mA)\\
\hline
Current consumption for receiving & 39.5 (mA)\\
\hline
\end{tabular}
\end{footnotesize}
\end{table}

\begin{figure}[htb]
\hspace{-0.80cm}\includegraphics[width=10.5cm,height=4.75cm]{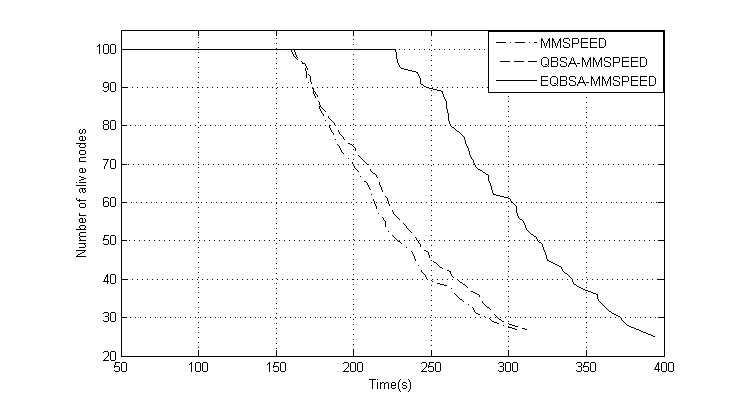}
\caption{Number of alive nodes at runtime}
\label{EC}
\end{figure}

Fig. \ref{EC} shows the average number of alive nodes during the video transmission over 50 realisations. We compare here three routing protocols: MMSPEED \cite{mmspeed}, QBSA-MMSPEED \cite{our} and the proposed EQBSA-MMSPEED. As can be seen, the proposed protocol extends the network lifetime of  about 33\%.

\begin{figure}[htb]
\centering \includegraphics[width=7cm,height=3.5cm]{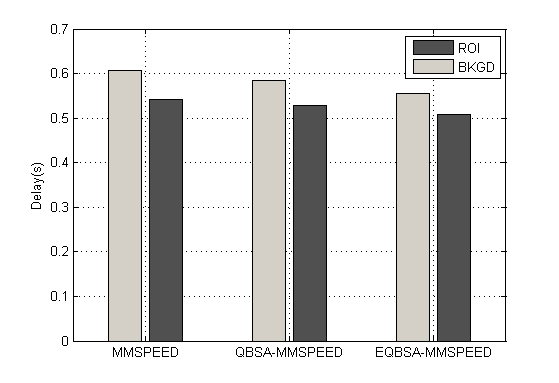}
\caption{Delays during the Rush mode}
\label{D_RM}
\end{figure}

Fig. \ref{D_RM} reports the results in the timeliness domain during the Rush mode. 
One can observe that the three protocols treat the ROI and the BKGD differently. In addition, we observe the slight enhancement in the delays that is achieved by the EQBSA-MMSPEED. This can be explained by the introduced load balance thanks to the consideration of the residual energy. Hence, nodes with appropriate delay and ABS are not always selected to not lead them to the state of congestion, thus having less packets to be processed.  

\begin{figure}[htb]
\centering \includegraphics[width=7cm,height=3.5cm]{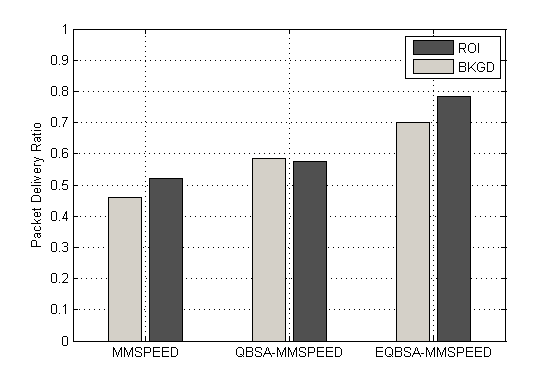}
\caption{Packet delivery ratio during the Rush mode}
\label{PDR_RM}
\end{figure}

Fig.\ref{PDR_RM} depicts the packet delivery ratio during the Rush mode.
We can observe that the proposed EQBSA-MMSPEED achieves interesting performances; this is due to four main reasons. The first one is, as said before, introducing new metric enables more load balancing. The second one is, even in worst cases, packet drop is avoided as much as possible while considering the energy consumption, thanks to the LCP. The third one is related to the fact that ignoring the remaining energy, in MMSPEED and QBSA-MMSPEED protocols, leads to a sort of acceleration in nodes' death. This fact creates holes specially in regions that are next to the destination, and this last becomes isolated. Consequently this leads to not receiving the packets at the sink. The last one is that the source nodes play also the role of forwarding nodes of others nodes' packets. Not considering their remaining energies accelerate their death as well, leading to their inability to send the whole video stream.
\vspace{-0.5cm}
\begin{figure}[htb]
\hspace{-0.80cm}\includegraphics[width=10.5cm,height=4.5cm]{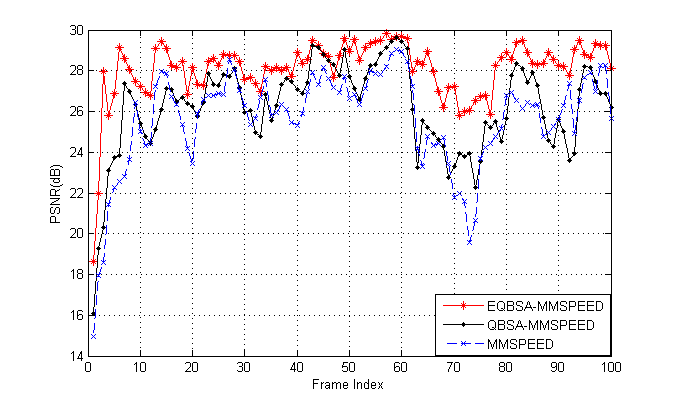}
\caption{PSNR values of the displayed stream}
\label{PSNR}
\end{figure}

Finally, Fig. \ref{PSNR} reports the performances of the proposed cross-layer approach with EQBSA-MMSPEED against MMSPEED \cite{mmspeed} and QBSA-MMSPEED \cite{our} in terms of video quality expressed by the Peak Signal to noise Ratio (PSNR). Note that, at the reception, a simple spatial error concealment procedure is applied to the received streams which exploits the inherent temporal correlation feature of the video signals. The proposed EQBSA-MMSPEED protocol allows the achievement of an average overall video quality enhancement of 3 dBs against MMSPEED and QBSA-MMSPEED.

\section{Conclusion}
\label{conclu}
In this paper, a cross-layer approach for video delivery in the WVSNs is proposed. The approach relies on a energy efficient and adaptive video compression scheme that has proven its efficiency in both video quality and energy domains. The scheme is based on the H.264/AVC standard and uses simple techniques in order to be energy efficient. Finally, it outputs two macroblock categories namely the region of interest and the background. These lasts are handled by an Energy and Queue Buffer Size Aware MMSPEED protocol that is able to offer differentiated service with the consideration of the available space in the queue buffer size of adjacent nodes, their remaining energy as well as their reliability. The proposed protocol achieves 33\% of lifetime extension and 3 dBs of video quality enhancement.

\bibliographystyle{IEEEtran}
\bibliography{refs}

%\begin{thebibliography}{1}


%\end{thebibliography}

\end{document}